\documentclass[a4paper,10pt,twocolumn]{revtex4-1}
\usepackage{graphicx}
\usepackage{amssymb}
\usepackage{amsmath} 
\usepackage{dcolumn} 
\usepackage{bm}  
\usepackage[english]{babel} 
\usepackage[novolumeabbr]{unitsdef}
\usepackage{fancyhdr}
\usepackage{color}
\usepackage[pdftex,                
bookmarks=true,
bookmarksnumbered=true,
hypertexnames=false,
breaklinks=true,
linkbordercolor={0 0 1}]{hyperref}

\newunit{\nanowatt}{\nano\watt}
\newunit{\sqmicm}{\micm^2}

\begin{document}
\title{Superconducting Nanowire Single Photon Detectors On-Fiber}
\author{Gil Bachar}
\email{gil@tx.technion.ac.il}
\author{Ilya Baskin}
\author{Oleg Shtempluck}
\author{Eyal Buks}
\affiliation{Department of Electrical Engineering, Technion, Haifa 32000, Israel}
\begin{abstract}
We present a novel design of a superconducting nanowire single photon detector (SNSPD) fabricated on a core of a single mode optical fiber. The proposed design allows high overlap between the fiber light mode and the detector, and consequently, our fabricated devices can remain small in dimension and maintain speed of operation, without scarifying the detection efficiency. The on-fiber fabrication method is detailed, together with experimental results. The proposed method can be exploited in the future for the fabrication of other fiber coupled devices.
\end{abstract}

\maketitle

Single photon detectors are needed in various fields of science and technology \cite{Natarajan_063001, hadfield_241129, Hadfield_10846, Gilbert_03, gobby_3762, Grein_78, Day_817, Baselmans_524, Alaverdian_2804}. The main figures of merit used to qualify a single photon detector are: the detection rate, the detection efficiency, the jitter, and the dark counts rate. Superconducting nanowire single photon detectors (SNSPD) \cite{Goltsman_705} are considered as a promising technology for an optical detection in the visible to near-infra-red band \cite{Natarajan_063001}. Achieving high coupling efficiency ($\eta_\mathrm{C}$) between the light source and the detector remains an outstanding challenge.

Maximizing $\eta_\mathrm{C}$ requires focusing the input light on a detector, which is typically defined as a square (or a circle) with few micrometers side (or diameter), and is operated inside a cryostat. Methods based on free space optics \cite{Goltsman_705} or mechanical cryogenic positioning of an optical fiber in front of the the sample \cite{Hu_3607} require complicated and expensive instruments, and suffer from poor alignment stability. Alternatively, methods based on fixed alignment of an optical fiber to the sample \cite{Miki_285, Slysz_261113} suffer from a fiber-center to detector-center misalignment (referred hereafter as the center-to-center-misalignment or $x_\mathrm{CC}$) of a few micrometers at least \cite{Miller_9102}. For these fixed alignment procedures, an increase in the detector dimensions is inevitable, in order to keep $\eta_\mathrm{C}$ high. As the detector recovery time is linearly proportional to its area \cite{Kerman_111116, Annunziata_084507}, it is important to keep the detector as small as possible. For example, for a typical device, made of $100\nm$ wide $5 \nm$ thick niobium-nitride wires folded to an area of $25\sqmicm$, the recovery time is $2.5\ns$ \cite{Kerman_111116, Annunziata_084507}

In the present work, we propose an alternative system configuration, in which the detector is fabricated on a tip of an optical fiber. The on-fiber fabrication allows precise alignment of the detector to the fiber core, where the light intensity is maximal, while keeping the device small and fast. The proposed devices are simple to operate, small in size and do not require complicated optical or positioning  equipment.

\begin{figure}
\includegraphics[width=\columnwidth]{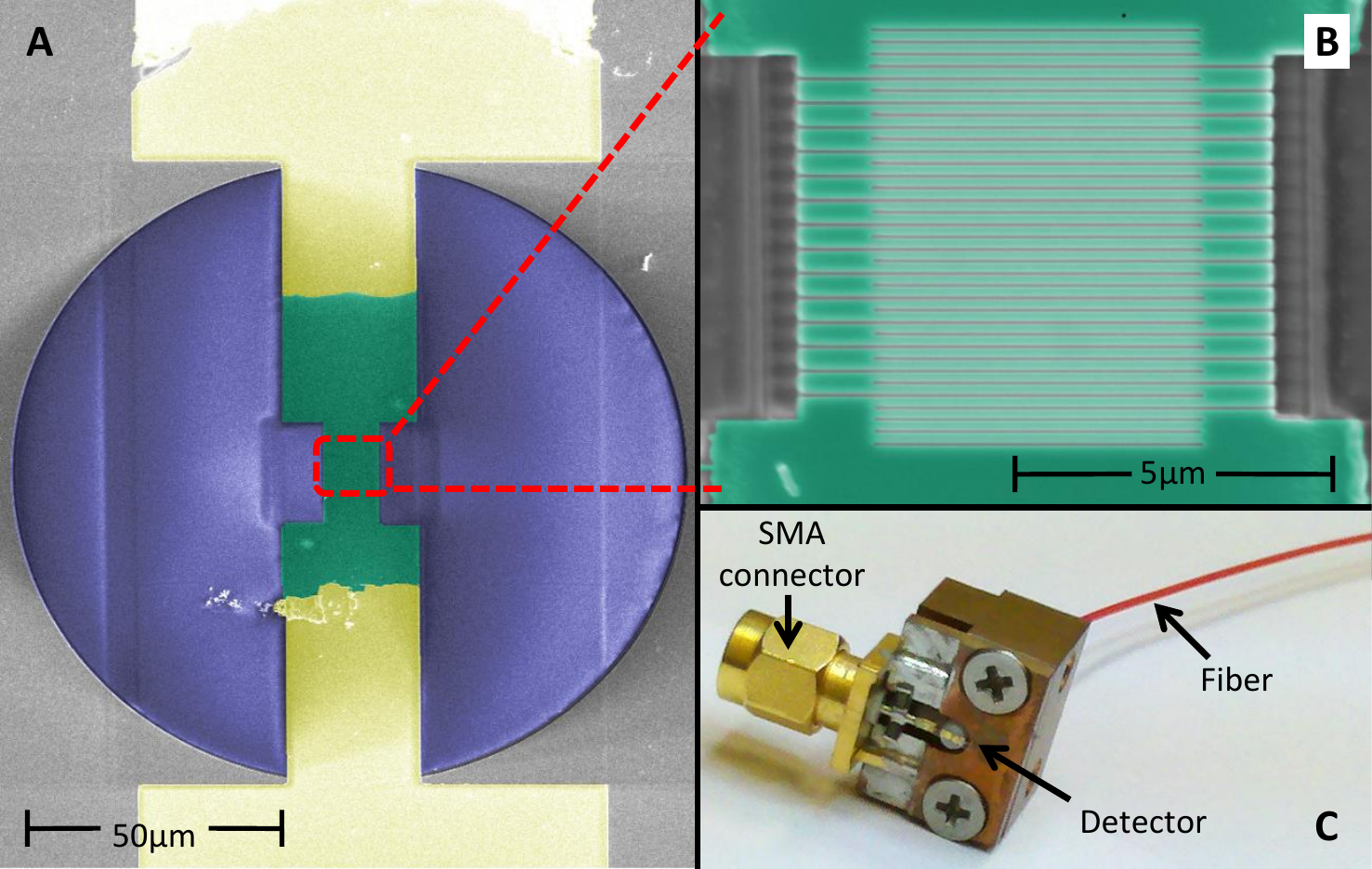}
\caption{Images of the detector. (A) A low magnification scanning electron microscope (SEM) micrograph of a SNSPD on a tip of a fiber (colored for clarity): the detector (green), the gold contacts (yellow) and the fiber (purple circle). (B) A higher magnification image. The thin $120\nm$ wide lines are folded in a meander form. The meander covers an area of $25\sqmicm$. (C) A photo of a mounted device.
}\label{fig:SEMImage}
\end{figure}

The complete fabrication process is done on the top facet of a zirconia ferrule, taken from a standard flat polished fiber connector (FC-UPC), holding a single mode fiber for the telecommunication bandwidth (Corning SMF-28 \cite{CorningWebsite}). The ferrule facet has a diameter of $2 \mm$; the fiber having a $125\micm$ diameter is epoxy glued concentrically with the ferrule.
We evaporate $5 \nm$ thick chromium film followed by a $200 \nm$ thick gold film trough a mechanical mask to form bonding pads.
Next, a $10 \nm$ thick niobium nitride (NbN) film is deposited using a dc-magnetron sputtering system. The sputtering process is done from a niobium target in a vacuum chamber filled with mixture of argon and nitrogen gasses, while keeping the sample at room temperature \cite{Marsili_3191, Bacon_6509}. The NbN film is covered in-situ with $100 \nm$ of aluminum.

To pattern the detector, we first narrow the NbN film to a $25\micm$ wide bridge using a focused ion beam (FIB) system with a relatively high current (2.1\nA). During this step the aluminum layer protects the NbN film from an exposure to the ion beam, and reduces gallium poisoning \cite{Tettamanzi_5302, Troeman_2152, Yuvaraj_arxiv2010}. After the first lithography step, the aluminum layer is wet etched. The sample then undergoes a second lithography step, in which a meander is formed, using a low current ($9.7 \micA$) FIB patterning. A fabricated device can be seen in Fig. \ref{fig:SEMImage}.

\begin{figure}
\includegraphics[width=\columnwidth]{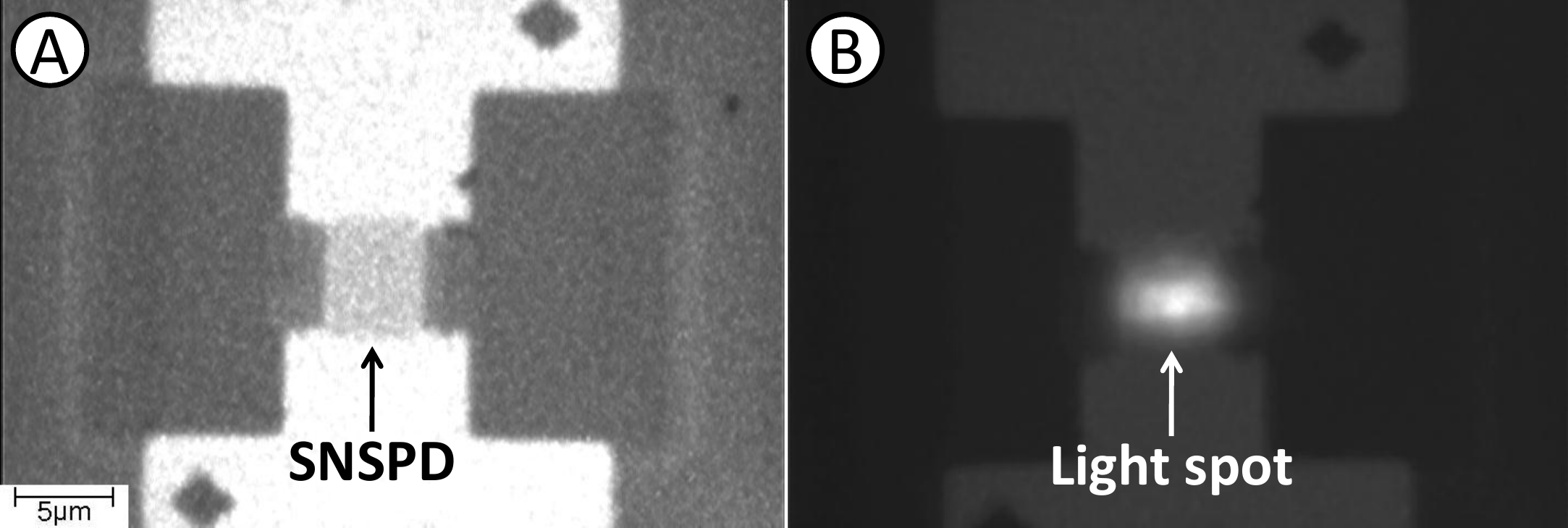}%
\caption{An optical image of a detector, obtained using IR sensitive camera. The nanowire is not visible, due to the optical resolution limitation (see Fig. \ref{fig:SEMImage}B for higher resolution). (A) the device under external illumination. (B) The external illumination is lowered and light is injected from the fiber, behind the detector (see Fig. \ref{fig:SEMImage}C). The high accuracy alignment of the light spot to the detector is demonstrated; the center of the light Gaussian beam overlaps the area of the detector.}\label{fig:OpticalImage}
\end{figure}

The FIB imaging system, which is used to align the detector to the fiber center, allows achieving $x_\mathrm{CC}$ of less than $1\micm$, limited mainly by the fiber-core to fiber-clad misalignment, given to be $<0.5\micm$ by Corning \cite{CorningWebsite}. In Fig. \ref{fig:OpticalImage} the low center-to-center-misalignment in our devices is demonstrated.

\begin{figure}
\includegraphics[width=\columnwidth]{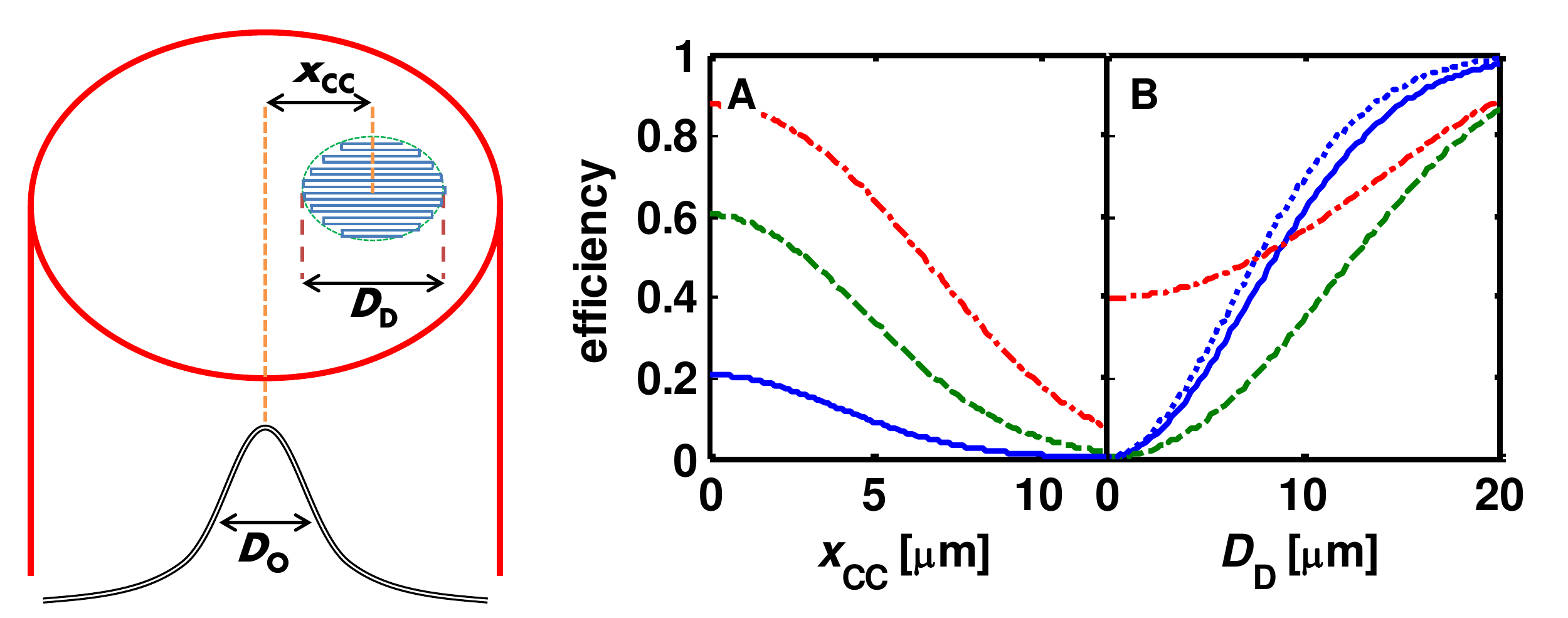}\\
\caption{The coupling efficiency ($\eta_\mathrm{C}$) between the Gaussian light beam with diameter $D_\mathrm{0}$ and the detector with size $D_\mathrm{D}$, where the center-center-misalignment is $x_\mathrm{CC}$. We assume $D_0 = 10.4 \micm$ (SMF-28), and detector fill factor of 1. (A) $\eta_\mathrm{C}$ vs. $x_\mathrm{CC}$ for circular detectors with $D_\mathrm{D} = 5$,$10$ and $15 \micm$ in solid-blue, dash-green and dash-dot-red respectively. (B) $\eta_\mathrm{C}$ vs. $D_\mathrm{D}$. In solid-blue (dotted-blue) $x_\mathrm{CC}=0$ for  a circular (rectangular) detector with diameter (edge) of $D_\mathrm{D}$. In dash-green, $x_\mathrm{CC}= 5\micm$. In dash-dot-red: the light loss due to $x_\mathrm{CC}= 5\micm$ relative to the $x_\mathrm{CC}=0$ case, i.e. the solid-blue curve divided by the dash-green curve.} \label{fig:OverlapFunctions}
\end{figure}

In order to estimate the effect of the center-center-misalignment on the detection efficiency, we calculate $\eta_\mathrm{C}$, namely the overlap between the Gaussian optical mode in an optical fiber \cite{YarivOptEl} and a detector located close to the top facet of the fiber:
\begin{equation}
\eta_\mathrm{C} = 4(\pi D_0^2)^{-1} \iint \exp (-4r^2/D_0^2) \times g(\textbf{r}) dA,
\end{equation}
where $D_0$ is the the mode field diameter and $g(\textbf{r})$ takes the a value of 1 if $\textbf{r}$ is in the area of the detector, and vanishes otherwise. In Fig. \ref{fig:OverlapFunctions} we plot $\eta_\mathrm{C}$ for a several detectors, with several $x_\mathrm{CC}$ values, where we assume detector area fill-factor of $100\%$ and a SMF-28 fiber. We can see that for $x_\mathrm{CC} \ge 10\micm$ and relatively large detector with diameter of $15\micm$ (panel A), $\eta_\mathrm{C}$ is less then $20\%$. Even for $x_\mathrm{CC} = 5\micm$ (panel B), $\eta_\mathrm{C}$ is limited by $50\%$ for detectors with less then $10\micm$ diameter. Note that for shorter wavelength fibers, the mode field diameter is even smaller 
and consequently misalignment results in even larger light loss.

\begin{figure}
\includegraphics[width=\columnwidth]{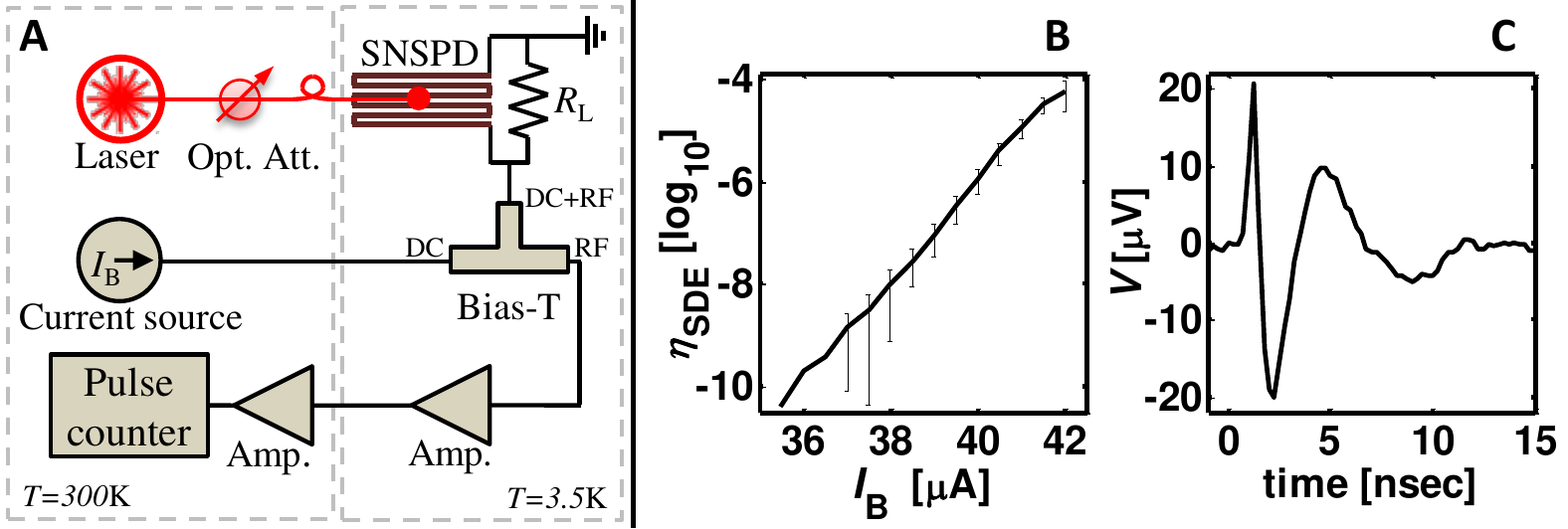}
\caption{(A) The experimental setup. The device is connected in parallel to a load resistor, $R_\mathrm{L}$, to prevent latching \cite{Annunziata_084507}. We use CW monochromatic laser at $1550\nm$ as a light source. The light is attenuated at room temperature to $10\nanowatt$ and sent into a cryostat where the device is cooled to $3.5\kelvin$. Current is sourced from a computer-controlled battery. The current enters the direct current (DC) port of a bias-T and into the detector. The output signal is sent from the RF port of the bias-T trough a cryogenic amplifier and up to room temperature using a semirigid coax cable. The signal is further amplified at room temperature and sent to a pulse counter (Stanford Research SR-400). (B) The system detection efficiency ($\eta_\mathrm{SDE}$) of one of our devices as a function of the bias current. The error-bars represent the standard deviation for the measurement. (C) An electrical pulse from the detector after an event of a single photon detection.}
\label{fig:ExpSetup}
\end{figure}

To characterize the detection performance, the fabricated device is inserted to a compact holder (Fig. \ref{fig:SEMImage}C), and is wire bonded to a SMA connector. The SMA is used for both DC current bias and fast pulses output. The experimental setup is schematically presented in Fig. \ref{fig:ExpSetup}A. In Fig. \ref{fig:ExpSetup}B, we show the system detection efficiency (noted hereafter as $\eta_\mathrm{SDE}$) as a function of the bias current in one of our devices. The critical current in the presented device is $I_\mathrm{C}=42\microampere$. The dark count, when the device is biased just below $I_\mathrm{C}$, is measured to be $10 \hertz$. Since $\eta_\mathrm{SDE}$ does not saturate when approaching $I_\mathrm{C}$ \cite{Natarajan_063001} and since the dark count rate is low compared to other devices reported in literature \cite{Natarajan_063001}, we conclude
that the level of the uniformity of critical current along the meander is relatively low in our device. Further work is needed to identify the underlying mechanisms that are responsible for the poor uniformity.

To summarize, we demonstrate a SNSPD fabricated on a tip of an optical fiber, with very low misalignment between the incoming light and the detector area. Though we achieve high coupling efficiency, more research is needed in order to increase the system detection efficiency. A future work in this direction may include adding an on-fiber optical cavity \cite{Baskin_arxiv}, changing the superconducting  material to a one more suitable for deposition on the amorphous $\mathrm{SiO_2}$ fiber \cite{Dorenbos_131101, Marsili_arxiv} and cooling the device to lower temperatures \cite{Engel_arxiv}.

This work was supported by the Mitchel Foundation, the Israel Ministry of Science, the Russell Berrie Nanotechnology Institute and MAFAT. The work of IB was supported by a Zeff Fellowship.

\bibliography{../../../Eyal_Bib}

\begin{thebibliography}{27}%
\makeatletter
\providecommand \@ifxundefined [1]{%
 \@ifx{#1\undefined}
}%
\providecommand \@ifnum [1]{%
 \ifnum #1\expandafter \@firstoftwo
 \else \expandafter \@secondoftwo
 \fi
}%
\providecommand \@ifx [1]{%
 \ifx #1\expandafter \@firstoftwo
 \else \expandafter \@secondoftwo
 \fi
}%
\providecommand \natexlab [1]{#1}%
\providecommand \enquote  [1]{``#1''}%
\providecommand \bibnamefont  [1]{#1}%
\providecommand \bibfnamefont [1]{#1}%
\providecommand \citenamefont [1]{#1}%
\providecommand \href@noop [0]{\@secondoftwo}%
\providecommand \href [0]{\begingroup \@sanitize@url \@href}%
\providecommand \@href[1]{\@@startlink{#1}\@@href}%
\providecommand \@@href[1]{\endgroup#1\@@endlink}%
\providecommand \@sanitize@url [0]{\catcode `\\12\catcode `\$12\catcode
  `\&12\catcode `\#12\catcode `\^12\catcode `\_12\catcode `\%12\relax}%
\providecommand \@@startlink[1]{}%
\providecommand \@@endlink[0]{}%
\providecommand \url  [0]{\begingroup\@sanitize@url \@url }%
\providecommand \@url [1]{\endgroup\@href {#1}{\urlprefix }}%
\providecommand \urlprefix  [0]{URL }%
\providecommand \Eprint [0]{\href }%
\providecommand \doibase [0]{http://dx.doi.org/}%
\providecommand \selectlanguage [0]{\@gobble}%
\providecommand \bibinfo  [0]{\@secondoftwo}%
\providecommand \bibfield  [0]{\@secondoftwo}%
\providecommand \translation [1]{[#1]}%
\providecommand \BibitemOpen [0]{}%
\providecommand \bibitemStop [0]{}%
\providecommand \bibitemNoStop [0]{.\EOS\space}%
\providecommand \EOS [0]{\spacefactor3000\relax}%
\providecommand \BibitemShut  [1]{\csname bibitem#1\endcsname}%
\let\auto@bib@innerbib\@empty
\bibitem [{\citenamefont {Natarajan}\ \emph {et~al.}(2012)\citenamefont
  {Natarajan}, \citenamefont {Tanner},\ and\ \citenamefont
  {Hadfield}}]{Natarajan_063001}%
  \BibitemOpen
  \bibfield  {author} {\bibinfo {author} {\bibfnamefont {C.~M.}\ \bibnamefont
  {Natarajan}}, \bibinfo {author} {\bibfnamefont {M.~G.}\ \bibnamefont
  {Tanner}}, \ and\ \bibinfo {author} {\bibfnamefont {R.~H.}\ \bibnamefont
  {Hadfield}},\ }\href {\doibase 10.1088/0953-2048/25/6/063001} {\bibfield
  {journal} {\bibinfo  {journal} {Supercond, Sci. and Tech.}\ }\textbf
  {\bibinfo {volume} {25}},\ \bibinfo {pages} {063001} (\bibinfo {year}
  {2012})}\BibitemShut {NoStop}%
\bibitem [{\citenamefont {Hadfield}\ \emph {et~al.}(2006)\citenamefont
  {Hadfield}, \citenamefont {Habif}, \citenamefont {Schlafer}, \citenamefont
  {Schwall},\ and\ \citenamefont {Nam}}]{hadfield_241129}%
  \BibitemOpen
  \bibfield  {author} {\bibinfo {author} {\bibfnamefont {R.~H.}\ \bibnamefont
  {Hadfield}}, \bibinfo {author} {\bibfnamefont {J.~L.}\ \bibnamefont {Habif}},
  \bibinfo {author} {\bibfnamefont {J.}~\bibnamefont {Schlafer}}, \bibinfo
  {author} {\bibfnamefont {R.~E.}\ \bibnamefont {Schwall}}, \ and\ \bibinfo
  {author} {\bibfnamefont {S.~W.}\ \bibnamefont {Nam}},\ }\href {\doibase
  10.1063/1.2405870} {\bibfield  {journal} {\bibinfo  {journal} {Appl. Phys.
  Lett.}\ }\textbf {\bibinfo {volume} {89}},\ \bibinfo {eid} {241129} (\bibinfo
  {year} {2006})}\BibitemShut {NoStop}%
\bibitem [{\citenamefont {Hadfield}\ \emph {et~al.}(2005)\citenamefont
  {Hadfield}, \citenamefont {Stevens}, \citenamefont {Gruber}, \citenamefont
  {Miller}, \citenamefont {Schwall}, \citenamefont {Mirin},\ and\ \citenamefont
  {Nam}}]{Hadfield_10846}%
  \BibitemOpen
  \bibfield  {author} {\bibinfo {author} {\bibfnamefont {R.~H.}\ \bibnamefont
  {Hadfield}}, \bibinfo {author} {\bibfnamefont {M.~J.}\ \bibnamefont
  {Stevens}}, \bibinfo {author} {\bibfnamefont {S.~S.}\ \bibnamefont {Gruber}},
  \bibinfo {author} {\bibfnamefont {A.~J.}\ \bibnamefont {Miller}}, \bibinfo
  {author} {\bibfnamefont {R.~E.}\ \bibnamefont {Schwall}}, \bibinfo {author}
  {\bibfnamefont {R.~P.}\ \bibnamefont {Mirin}}, \ and\ \bibinfo {author}
  {\bibfnamefont {S.~W.}\ \bibnamefont {Nam}},\ }\href {\doibase
  10.1364/OPEX.13.010846} {\bibfield  {journal} {\bibinfo  {journal} {Opt.
  Express}\ }\textbf {\bibinfo {volume} {13}},\ \bibinfo {pages} {10846}
  (\bibinfo {year} {2005})}\BibitemShut {NoStop}%
\bibitem [{\citenamefont {Gilbert}\ and\ \citenamefont
  {Hamrick}(2000)}]{Gilbert_03}%
  \BibitemOpen
  \bibfield  {author} {\bibinfo {author} {\bibfnamefont {G.}~\bibnamefont
  {Gilbert}}\ and\ \bibinfo {author} {\bibfnamefont {M.}~\bibnamefont
  {Hamrick}},\ }\href@noop {} {\bibfield  {journal} {\bibinfo  {journal}
  {arXiv:quant-ph/0009027}\ } (\bibinfo {year} {2000})}\BibitemShut {NoStop}%
\bibitem [{\citenamefont {Gobby}\ \emph {et~al.}(2004)\citenamefont {Gobby},
  \citenamefont {Yuan},\ and\ \citenamefont {Shields}}]{gobby_3762}%
  \BibitemOpen
  \bibfield  {author} {\bibinfo {author} {\bibfnamefont {C.}~\bibnamefont
  {Gobby}}, \bibinfo {author} {\bibfnamefont {Z.~L.}\ \bibnamefont {Yuan}}, \
  and\ \bibinfo {author} {\bibfnamefont {A.~J.}\ \bibnamefont {Shields}},\
  }\href {\doibase 10.1063/1.1738173} {\bibfield  {journal} {\bibinfo
  {journal} {Appl. Phys. Lett.}\ }\textbf {\bibinfo {volume} {84}},\ \bibinfo
  {pages} {3762} (\bibinfo {year} {2004})}\BibitemShut {NoStop}%
\bibitem [{\citenamefont {Grein}\ \emph {et~al.}(2011)\citenamefont {Grein},
  \citenamefont {Kerman}, \citenamefont {Dauler}, \citenamefont {Shatrovoy},
  \citenamefont {Molnar}, \citenamefont {Rosenberg}, \citenamefont {Yoon},
  \citenamefont {DeVoe}, \citenamefont {Murphy}, \citenamefont {Robinson},\
  and\ \citenamefont {Boroson}}]{Grein_78}%
  \BibitemOpen
  \bibfield  {author} {\bibinfo {author} {\bibfnamefont {M.}~\bibnamefont
  {Grein}}, \bibinfo {author} {\bibfnamefont {A.}~\bibnamefont {Kerman}},
  \bibinfo {author} {\bibfnamefont {E.}~\bibnamefont {Dauler}}, \bibinfo
  {author} {\bibfnamefont {O.}~\bibnamefont {Shatrovoy}}, \bibinfo {author}
  {\bibfnamefont {R.}~\bibnamefont {Molnar}}, \bibinfo {author} {\bibfnamefont
  {D.}~\bibnamefont {Rosenberg}}, \bibinfo {author} {\bibfnamefont
  {J.}~\bibnamefont {Yoon}}, \bibinfo {author} {\bibfnamefont {C.}~\bibnamefont
  {DeVoe}}, \bibinfo {author} {\bibfnamefont {D.}~\bibnamefont {Murphy}},
  \bibinfo {author} {\bibfnamefont {B.}~\bibnamefont {Robinson}}, \ and\
  \bibinfo {author} {\bibfnamefont {D.}~\bibnamefont {Boroson}},\ }in\ \href
  {\doibase 10.1109/ICSOS.2011.5783715} {\emph {\bibinfo {booktitle} {Space
  Optical Systems and Applications (ICSOS), 2011 International Conference
  on}}}\ (\bibinfo {year} {2011})\ pp.\ \bibinfo {pages} {78 --82}\BibitemShut
  {NoStop}%
\bibitem [{\citenamefont {Day}\ \emph {et~al.}(2003)\citenamefont {Day},
  \citenamefont {LeDuc}, \citenamefont {Mazin}, \citenamefont {Vayonakis},\
  and\ \citenamefont {Zmuidzinas}}]{Day_817}%
  \BibitemOpen
  \bibfield  {author} {\bibinfo {author} {\bibfnamefont {P.~K.}\ \bibnamefont
  {Day}}, \bibinfo {author} {\bibfnamefont {H.~G.}\ \bibnamefont {LeDuc}},
  \bibinfo {author} {\bibfnamefont {B.~A.}\ \bibnamefont {Mazin}}, \bibinfo
  {author} {\bibfnamefont {A.}~\bibnamefont {Vayonakis}}, \ and\ \bibinfo
  {author} {\bibfnamefont {J.}~\bibnamefont {Zmuidzinas}},\ }\href {\doibase
  10.1038/nature02037} {\bibfield  {journal} {\bibinfo  {journal} {Nature}\
  }\textbf {\bibinfo {volume} {425}},\ \bibinfo {pages} {817} (\bibinfo {year}
  {2003})}\BibitemShut {NoStop}%
\bibitem [{\citenamefont {{Baselmans}}\ \emph {et~al.}(2008)\citenamefont
  {{Baselmans}}, \citenamefont {{Yates}}, \citenamefont {{Barends}},
  \citenamefont {{Lankwarden}}, \citenamefont {{Gao}}, \citenamefont
  {{Hoevers}},\ and\ \citenamefont {{Klapwijk}}}]{Baselmans_524}%
  \BibitemOpen
  \bibfield  {author} {\bibinfo {author} {\bibfnamefont {J.}~\bibnamefont
  {{Baselmans}}}, \bibinfo {author} {\bibfnamefont {S.}~\bibnamefont
  {{Yates}}}, \bibinfo {author} {\bibfnamefont {R.}~\bibnamefont {{Barends}}},
  \bibinfo {author} {\bibfnamefont {Y.}~\bibnamefont {{Lankwarden}}}, \bibinfo
  {author} {\bibfnamefont {J.}~\bibnamefont {{Gao}}}, \bibinfo {author}
  {\bibfnamefont {H.}~\bibnamefont {{Hoevers}}}, \ and\ \bibinfo {author}
  {\bibfnamefont {T.}~\bibnamefont {{Klapwijk}}},\ }\href {\doibase
  10.1007/s10909-007-9684-3} {\bibfield  {journal} {\bibinfo  {journal} {J. Low
  Temp. Phys}\ }\textbf {\bibinfo {volume} {151}},\ \bibinfo {pages} {524}
  (\bibinfo {year} {2008})}\BibitemShut {NoStop}%
\bibitem [{\citenamefont {Alaverdian}\ \emph {et~al.}(2002)\citenamefont
  {Alaverdian}, \citenamefont {Alaverdian}, \citenamefont {Bilenko},
  \citenamefont {Bogdanov}, \citenamefont {Filippova}, \citenamefont
  {Gavrilov}, \citenamefont {Gorbovitski}, \citenamefont {Gouzman},
  \citenamefont {Gudkov}, \citenamefont {Domratchev}, \citenamefont
  {Kosobokova}, \citenamefont {Lifshitz}, \citenamefont {Luryi}, \citenamefont
  {Ruskovoloshin}, \citenamefont {Stepoukhovitch}, \citenamefont
  {Tcherevishnick}, \citenamefont {Tyshko},\ and\ \citenamefont
  {Gorfinkel}}]{Alaverdian_2804}%
  \BibitemOpen
  \bibfield  {author} {\bibinfo {author} {\bibfnamefont {L.}~\bibnamefont
  {Alaverdian}}, \bibinfo {author} {\bibfnamefont {S.}~\bibnamefont
  {Alaverdian}}, \bibinfo {author} {\bibfnamefont {O.}~\bibnamefont {Bilenko}},
  \bibinfo {author} {\bibfnamefont {I.}~\bibnamefont {Bogdanov}}, \bibinfo
  {author} {\bibfnamefont {E.}~\bibnamefont {Filippova}}, \bibinfo {author}
  {\bibfnamefont {D.}~\bibnamefont {Gavrilov}}, \bibinfo {author}
  {\bibfnamefont {B.}~\bibnamefont {Gorbovitski}}, \bibinfo {author}
  {\bibfnamefont {M.}~\bibnamefont {Gouzman}}, \bibinfo {author} {\bibfnamefont
  {G.}~\bibnamefont {Gudkov}}, \bibinfo {author} {\bibfnamefont
  {S.}~\bibnamefont {Domratchev}}, \bibinfo {author} {\bibfnamefont
  {O.}~\bibnamefont {Kosobokova}}, \bibinfo {author} {\bibfnamefont
  {N.}~\bibnamefont {Lifshitz}}, \bibinfo {author} {\bibfnamefont
  {S.}~\bibnamefont {Luryi}}, \bibinfo {author} {\bibfnamefont
  {V.}~\bibnamefont {Ruskovoloshin}}, \bibinfo {author} {\bibfnamefont
  {A.}~\bibnamefont {Stepoukhovitch}}, \bibinfo {author} {\bibfnamefont
  {M.}~\bibnamefont {Tcherevishnick}}, \bibinfo {author} {\bibfnamefont
  {G.}~\bibnamefont {Tyshko}}, \ and\ \bibinfo {author} {\bibfnamefont
  {V.}~\bibnamefont {Gorfinkel}},\ }\href {\doibase
  10.1002/1522-2683(200208)23:16<2804::AID-ELPS2804>3.0.CO;2-9} {\bibfield
  {journal} {\bibinfo  {journal} {Electrophoresis}\ }\textbf {\bibinfo {volume}
  {23}},\ \bibinfo {pages} {2804} (\bibinfo {year} {2002})}\BibitemShut
  {NoStop}%
\bibitem [{\citenamefont {Gol'tsman}\ \emph {et~al.}(2001)\citenamefont
  {Gol'tsman}, \citenamefont {Okunev}, \citenamefont {Chulkova}, \citenamefont
  {Lipatov}, \citenamefont {Semenov}, \citenamefont {Smirnov}, \citenamefont
  {Voronov}, \citenamefont {Dzardanov}, \citenamefont {Williams},\ and\
  \citenamefont {Sobolewski}}]{Goltsman_705}%
  \BibitemOpen
  \bibfield  {author} {\bibinfo {author} {\bibfnamefont {G.~N.}\ \bibnamefont
  {Gol'tsman}}, \bibinfo {author} {\bibfnamefont {O.}~\bibnamefont {Okunev}},
  \bibinfo {author} {\bibfnamefont {G.}~\bibnamefont {Chulkova}}, \bibinfo
  {author} {\bibfnamefont {A.}~\bibnamefont {Lipatov}}, \bibinfo {author}
  {\bibfnamefont {A.}~\bibnamefont {Semenov}}, \bibinfo {author} {\bibfnamefont
  {K.}~\bibnamefont {Smirnov}}, \bibinfo {author} {\bibfnamefont
  {B.}~\bibnamefont {Voronov}}, \bibinfo {author} {\bibfnamefont
  {A.}~\bibnamefont {Dzardanov}}, \bibinfo {author} {\bibfnamefont
  {C.}~\bibnamefont {Williams}}, \ and\ \bibinfo {author} {\bibfnamefont
  {R.}~\bibnamefont {Sobolewski}},\ }\href {\doibase 10.1063/1.1388868}
  {\bibfield  {journal} {\bibinfo  {journal} {Appl. Phys. Lett.}\ }\textbf
  {\bibinfo {volume} {79}},\ \bibinfo {pages} {705} (\bibinfo {year}
  {2001})}\BibitemShut {NoStop}%
\bibitem [{\citenamefont {Hu}\ \emph {et~al.}(2009)\citenamefont {Hu},
  \citenamefont {Zhong}, \citenamefont {White}, \citenamefont {Dauler},
  \citenamefont {Najafi}, \citenamefont {Herder}, \citenamefont {Wong},\ and\
  \citenamefont {Berggren}}]{Hu_3607}%
  \BibitemOpen
  \bibfield  {author} {\bibinfo {author} {\bibfnamefont {X.}~\bibnamefont
  {Hu}}, \bibinfo {author} {\bibfnamefont {T.}~\bibnamefont {Zhong}}, \bibinfo
  {author} {\bibfnamefont {J.~E.}\ \bibnamefont {White}}, \bibinfo {author}
  {\bibfnamefont {E.~A.}\ \bibnamefont {Dauler}}, \bibinfo {author}
  {\bibfnamefont {F.}~\bibnamefont {Najafi}}, \bibinfo {author} {\bibfnamefont
  {C.~H.}\ \bibnamefont {Herder}}, \bibinfo {author} {\bibfnamefont {F.~N.~C.}\
  \bibnamefont {Wong}}, \ and\ \bibinfo {author} {\bibfnamefont {K.~K.}\
  \bibnamefont {Berggren}},\ }\href {\doibase 10.1364/OL.34.003607} {\bibfield
  {journal} {\bibinfo  {journal} {Opt. Lett.}\ }\textbf {\bibinfo {volume}
  {34}},\ \bibinfo {pages} {3607} (\bibinfo {year} {2009})}\BibitemShut
  {NoStop}%
\bibitem [{\citenamefont {Miki}\ \emph {et~al.}(2007)\citenamefont {Miki},
  \citenamefont {Fujiwara}, \citenamefont {Sasaki},\ and\ \citenamefont
  {Wang}}]{Miki_285}%
  \BibitemOpen
  \bibfield  {author} {\bibinfo {author} {\bibfnamefont {S.}~\bibnamefont
  {Miki}}, \bibinfo {author} {\bibfnamefont {M.}~\bibnamefont {Fujiwara}},
  \bibinfo {author} {\bibfnamefont {M.}~\bibnamefont {Sasaki}}, \ and\ \bibinfo
  {author} {\bibfnamefont {Z.}~\bibnamefont {Wang}},\ }\href {\doibase
  10.1109/TASC.2007.898582} {\bibfield  {journal} {\bibinfo  {journal} {{{IEEE}
  Trans. Appl. Superconduct.}}\ }\textbf {\bibinfo {volume} {17}},\ \bibinfo
  {pages} {285 } (\bibinfo {year} {2007})}\BibitemShut {NoStop}%
\bibitem [{\citenamefont {Slysz}\ \emph {et~al.}(2006)\citenamefont {Slysz},
  \citenamefont {Wegrzecki}, \citenamefont {Bar}, \citenamefont {Grabiec},
  \citenamefont {G\'{o}rska}, \citenamefont {Zwiller}, \citenamefont {Latta},
  \citenamefont {Bohi}, \citenamefont {Milostnaya}, \citenamefont {Minaeva},
  \citenamefont {Antipov}, \citenamefont {Okunev}, \citenamefont {Korneev},
  \citenamefont {Smirnov}, \citenamefont {Voronov}, \citenamefont {Kaurova},
  \citenamefont {Gol'tsman}, \citenamefont {Pearlman}, \citenamefont {Cross},
  \citenamefont {Komissarov}, \citenamefont {Verevkin},\ and\ \citenamefont
  {Sobolewski}}]{Slysz_261113}%
  \BibitemOpen
  \bibfield  {author} {\bibinfo {author} {\bibfnamefont {W.}~\bibnamefont
  {Slysz}}, \bibinfo {author} {\bibfnamefont {M.}~\bibnamefont {Wegrzecki}},
  \bibinfo {author} {\bibfnamefont {J.}~\bibnamefont {Bar}}, \bibinfo {author}
  {\bibfnamefont {P.}~\bibnamefont {Grabiec}}, \bibinfo {author} {\bibfnamefont
  {M.}~\bibnamefont {G\'{o}rska}}, \bibinfo {author} {\bibfnamefont
  {V.}~\bibnamefont {Zwiller}}, \bibinfo {author} {\bibfnamefont
  {C.}~\bibnamefont {Latta}}, \bibinfo {author} {\bibfnamefont
  {P.}~\bibnamefont {Bohi}}, \bibinfo {author} {\bibfnamefont {I.}~\bibnamefont
  {Milostnaya}}, \bibinfo {author} {\bibfnamefont {O.}~\bibnamefont {Minaeva}},
  \bibinfo {author} {\bibfnamefont {A.}~\bibnamefont {Antipov}}, \bibinfo
  {author} {\bibfnamefont {O.}~\bibnamefont {Okunev}}, \bibinfo {author}
  {\bibfnamefont {A.}~\bibnamefont {Korneev}}, \bibinfo {author} {\bibfnamefont
  {K.}~\bibnamefont {Smirnov}}, \bibinfo {author} {\bibfnamefont
  {B.}~\bibnamefont {Voronov}}, \bibinfo {author} {\bibfnamefont
  {N.}~\bibnamefont {Kaurova}}, \bibinfo {author} {\bibfnamefont
  {G.}~\bibnamefont {Gol'tsman}}, \bibinfo {author} {\bibfnamefont
  {A.}~\bibnamefont {Pearlman}}, \bibinfo {author} {\bibfnamefont
  {A.}~\bibnamefont {Cross}}, \bibinfo {author} {\bibfnamefont
  {I.}~\bibnamefont {Komissarov}}, \bibinfo {author} {\bibfnamefont
  {A.}~\bibnamefont {Verevkin}}, \ and\ \bibinfo {author} {\bibfnamefont
  {R.}~\bibnamefont {Sobolewski}},\ }\href {\doibase 10.1063/1.2218105}
  {\bibfield  {journal} {\bibinfo  {journal} {Appl. Phys. Lett.}\ }\textbf
  {\bibinfo {volume} {88}},\ \bibinfo {eid} {261113} (\bibinfo {year}
  {2006})}\BibitemShut {NoStop}%
\bibitem [{\citenamefont {Miller}\ \emph {et~al.}(2011)\citenamefont {Miller},
  \citenamefont {Lita}, \citenamefont {Calkins}, \citenamefont {Vayshenker},
  \citenamefont {Gruber},\ and\ \citenamefont {Nam}}]{Miller_9102}%
  \BibitemOpen
  \bibfield  {author} {\bibinfo {author} {\bibfnamefont {A.~J.}\ \bibnamefont
  {Miller}}, \bibinfo {author} {\bibfnamefont {A.~E.}\ \bibnamefont {Lita}},
  \bibinfo {author} {\bibfnamefont {B.}~\bibnamefont {Calkins}}, \bibinfo
  {author} {\bibfnamefont {I.}~\bibnamefont {Vayshenker}}, \bibinfo {author}
  {\bibfnamefont {S.~M.}\ \bibnamefont {Gruber}}, \ and\ \bibinfo {author}
  {\bibfnamefont {S.~W.}\ \bibnamefont {Nam}},\ }\href {\doibase
  10.1364/OE.19.009102} {\bibfield  {journal} {\bibinfo  {journal} {Opt.
  Express}\ }\textbf {\bibinfo {volume} {19}},\ \bibinfo {pages} {9102}
  (\bibinfo {year} {2011})}\BibitemShut {NoStop}%
\bibitem [{\citenamefont {Kerman}\ \emph {et~al.}(2006)\citenamefont {Kerman},
  \citenamefont {Dauler}, \citenamefont {Keicher}, \citenamefont {Yang},
  \citenamefont {Berggren}, \citenamefont {Gol'tsman},\ and\ \citenamefont
  {Voronov}}]{Kerman_111116}%
  \BibitemOpen
  \bibfield  {author} {\bibinfo {author} {\bibfnamefont {A.~J.}\ \bibnamefont
  {Kerman}}, \bibinfo {author} {\bibfnamefont {E.~A.}\ \bibnamefont {Dauler}},
  \bibinfo {author} {\bibfnamefont {W.~E.}\ \bibnamefont {Keicher}}, \bibinfo
  {author} {\bibfnamefont {J.~K.~W.}\ \bibnamefont {Yang}}, \bibinfo {author}
  {\bibfnamefont {K.~K.}\ \bibnamefont {Berggren}}, \bibinfo {author}
  {\bibfnamefont {G.}~\bibnamefont {Gol'tsman}}, \ and\ \bibinfo {author}
  {\bibfnamefont {B.}~\bibnamefont {Voronov}},\ }\href {\doibase
  10.1063/1.2183810} {\bibfield  {journal} {\bibinfo  {journal} {Appl. Phys.
  Lett.}\ }\textbf {\bibinfo {volume} {88}},\ \bibinfo {eid} {111116} (\bibinfo
  {year} {2006})}\BibitemShut {NoStop}%
\bibitem [{\citenamefont {Annunziata}\ \emph {et~al.}(2010)\citenamefont
  {Annunziata}, \citenamefont {Quaranta}, \citenamefont {Santavicca},
  \citenamefont {Casaburi}, \citenamefont {Frunzio}, \citenamefont {Ejrnaes},
  \citenamefont {Rooks}, \citenamefont {Cristiano}, \citenamefont {Pagano},
  \citenamefont {Frydman},\ and\ \citenamefont {Prober}}]{Annunziata_084507}%
  \BibitemOpen
  \bibfield  {author} {\bibinfo {author} {\bibfnamefont {A.~J.}\ \bibnamefont
  {Annunziata}}, \bibinfo {author} {\bibfnamefont {O.}~\bibnamefont
  {Quaranta}}, \bibinfo {author} {\bibfnamefont {D.~F.}\ \bibnamefont
  {Santavicca}}, \bibinfo {author} {\bibfnamefont {A.}~\bibnamefont
  {Casaburi}}, \bibinfo {author} {\bibfnamefont {L.}~\bibnamefont {Frunzio}},
  \bibinfo {author} {\bibfnamefont {M.}~\bibnamefont {Ejrnaes}}, \bibinfo
  {author} {\bibfnamefont {M.~J.}\ \bibnamefont {Rooks}}, \bibinfo {author}
  {\bibfnamefont {R.}~\bibnamefont {Cristiano}}, \bibinfo {author}
  {\bibfnamefont {S.}~\bibnamefont {Pagano}}, \bibinfo {author} {\bibfnamefont
  {A.}~\bibnamefont {Frydman}}, \ and\ \bibinfo {author} {\bibfnamefont
  {D.~E.}\ \bibnamefont {Prober}},\ }\href {\doibase 10.1063/1.3498809}
  {\bibfield  {journal} {\bibinfo  {journal} {J. Appl. Phys.}\ }\textbf
  {\bibinfo {volume} {108}},\ \bibinfo {pages} {084507} (\bibinfo {year}
  {2010})}\BibitemShut {NoStop}%
\bibitem [{Cor()}]{CorningWebsite}%
  \BibitemOpen
  \href@noop {} {}\bibinfo {howpublished} {{Corning Inc. wbsite}
  \url{http://www.corning.com}}\BibitemShut {NoStop}%
\bibitem [{\citenamefont {Marsili}\ \emph {et~al.}(2008)\citenamefont
  {Marsili}, \citenamefont {Bitauld}, \citenamefont {Fiore}, \citenamefont
  {Gaggero}, \citenamefont {Mattioli}, \citenamefont {Leoni}, \citenamefont
  {Benkahoul},\ and\ \citenamefont {L\'{e}vy}}]{Marsili_3191}%
  \BibitemOpen
  \bibfield  {author} {\bibinfo {author} {\bibfnamefont {F.}~\bibnamefont
  {Marsili}}, \bibinfo {author} {\bibfnamefont {D.}~\bibnamefont {Bitauld}},
  \bibinfo {author} {\bibfnamefont {A.}~\bibnamefont {Fiore}}, \bibinfo
  {author} {\bibfnamefont {A.}~\bibnamefont {Gaggero}}, \bibinfo {author}
  {\bibfnamefont {F.}~\bibnamefont {Mattioli}}, \bibinfo {author}
  {\bibfnamefont {R.}~\bibnamefont {Leoni}}, \bibinfo {author} {\bibfnamefont
  {M.}~\bibnamefont {Benkahoul}}, \ and\ \bibinfo {author} {\bibfnamefont
  {F.}~\bibnamefont {L\'{e}vy}},\ }\href {\doibase 10.1364/OE.16.003191}
  {\bibfield  {journal} {\bibinfo  {journal} {Opt. Express}\ }\textbf {\bibinfo
  {volume} {16}},\ \bibinfo {pages} {3191} (\bibinfo {year}
  {2008})}\BibitemShut {NoStop}%
\bibitem [{\citenamefont {Bacon}\ \emph {et~al.}(1983)\citenamefont {Bacon},
  \citenamefont {{English}}, \citenamefont {{Nakahara}}, \citenamefont
  {{Peters}}, \citenamefont {{Schreiber}}, \citenamefont {{Sinclair}},\ and\
  \citenamefont {{van Dover}}}]{Bacon_6509}%
  \BibitemOpen
  \bibfield  {author} {\bibinfo {author} {\bibfnamefont {D.~D.}\ \bibnamefont
  {Bacon}}, \bibinfo {author} {\bibfnamefont {A.~T.}\ \bibnamefont
  {{English}}}, \bibinfo {author} {\bibfnamefont {S.}~\bibnamefont
  {{Nakahara}}}, \bibinfo {author} {\bibfnamefont {F.~G.}\ \bibnamefont
  {{Peters}}}, \bibinfo {author} {\bibfnamefont {H.}~\bibnamefont
  {{Schreiber}}}, \bibinfo {author} {\bibfnamefont {W.~R.}\ \bibnamefont
  {{Sinclair}}}, \ and\ \bibinfo {author} {\bibfnamefont {R.~B.}\ \bibnamefont
  {{van Dover}}},\ }\href {\doibase 10.1063/1.331881} {\bibfield  {journal}
  {\bibinfo  {journal} {J. Appl. Phys.}\ }\textbf {\bibinfo {volume} {54}},\
  \bibinfo {pages} {6509} (\bibinfo {year} {1983})}\BibitemShut {NoStop}%
\bibitem [{\citenamefont {{Tettamanzi}}\ \emph {et~al.}(2009)\citenamefont
  {{Tettamanzi}}, \citenamefont {{Pakes}}, \citenamefont {{Potenza}},
  \citenamefont {{Rubanov}}, \citenamefont {{Marrows}},\ and\ \citenamefont
  {{Prawer}}}]{Tettamanzi_5302}%
  \BibitemOpen
  \bibfield  {author} {\bibinfo {author} {\bibfnamefont {G.~C.}\ \bibnamefont
  {{Tettamanzi}}}, \bibinfo {author} {\bibfnamefont {C.~I.}\ \bibnamefont
  {{Pakes}}}, \bibinfo {author} {\bibfnamefont {A.}~\bibnamefont {{Potenza}}},
  \bibinfo {author} {\bibfnamefont {S.}~\bibnamefont {{Rubanov}}}, \bibinfo
  {author} {\bibfnamefont {C.~H.}\ \bibnamefont {{Marrows}}}, \ and\ \bibinfo
  {author} {\bibfnamefont {S.}~\bibnamefont {{Prawer}}},\ }\href {\doibase
  10.1088/0957-4484/20/46/465302} {\bibfield  {journal} {\bibinfo  {journal}
  {Nanotechnology}\ }\textbf {\bibinfo {volume} {20}},\ \bibinfo {pages} {5302}
  (\bibinfo {year} {2009})}\BibitemShut {NoStop}%
\bibitem [{\citenamefont {Troeman}\ \emph {et~al.}(2007)\citenamefont
  {Troeman}, \citenamefont {Derking}, \citenamefont {Borger}, \citenamefont
  {Pleikies}, \citenamefont {Veldhuis},\ and\ \citenamefont
  {Hilgenkamp}}]{Troeman_2152}%
  \BibitemOpen
  \bibfield  {author} {\bibinfo {author} {\bibfnamefont {A.}~\bibnamefont
  {Troeman}}, \bibinfo {author} {\bibfnamefont {H.}~\bibnamefont {Derking}},
  \bibinfo {author} {\bibfnamefont {B.}~\bibnamefont {Borger}}, \bibinfo
  {author} {\bibfnamefont {J.}~\bibnamefont {Pleikies}}, \bibinfo {author}
  {\bibfnamefont {D.}~\bibnamefont {Veldhuis}}, \ and\ \bibinfo {author}
  {\bibfnamefont {H.}~\bibnamefont {Hilgenkamp}},\ }\href {\doibase
  10.1021/nl070870f} {\bibfield  {journal} {\bibinfo  {journal} {Nano Lett.}\
  }\textbf {\bibinfo {volume} {7}},\ \bibinfo {pages} {2152} (\bibinfo {year}
  {2007})}\BibitemShut {NoStop}%
\bibitem [{\citenamefont {{Yuvaraj}}\ \emph {et~al.}(2011)\citenamefont
  {{Yuvaraj}}, \citenamefont {{Bachar}}, \citenamefont {{Suchoi}},
  \citenamefont {{Shtempluck}},\ and\ \citenamefont
  {{Buks}}}]{Yuvaraj_arxiv2010}%
  \BibitemOpen
  \bibfield  {author} {\bibinfo {author} {\bibfnamefont {D.}~\bibnamefont
  {{Yuvaraj}}}, \bibinfo {author} {\bibfnamefont {G.}~\bibnamefont {{Bachar}}},
  \bibinfo {author} {\bibfnamefont {O.}~\bibnamefont {{Suchoi}}}, \bibinfo
  {author} {\bibfnamefont {O.}~\bibnamefont {{Shtempluck}}}, \ and\ \bibinfo
  {author} {\bibfnamefont {E.}~\bibnamefont {{Buks}}},\ }\href@noop {}
  {\bibfield  {journal} {\bibinfo  {journal} {arXiv:1107.0635}\ } (\bibinfo
  {year} {2011})}\BibitemShut {NoStop}%
\bibitem [{\citenamefont {Yariv}(1985)}]{YarivOptEl}%
  \BibitemOpen
  \bibfield  {author} {\bibinfo {author} {\bibfnamefont {A.}~\bibnamefont
  {Yariv}},\ }\href@noop {} {\emph {\bibinfo {title} {Optical electronics}}}\
  (\bibinfo  {publisher} {Holt, Rinehart and Winston},\ \bibinfo {year}
  {1985})\BibitemShut {NoStop}%
\bibitem [{\citenamefont {Baskin}\ \emph {et~al.}(2012)\citenamefont {Baskin},
  \citenamefont {Bachar}, \citenamefont {Shlomi}, \citenamefont {Shtempluck},\
  and\ \citenamefont {Buks}}]{Baskin_arxiv}%
  \BibitemOpen
  \bibfield  {author} {\bibinfo {author} {\bibfnamefont {I.}~\bibnamefont
  {Baskin}}, \bibinfo {author} {\bibfnamefont {G.}~\bibnamefont {Bachar}},
  \bibinfo {author} {\bibfnamefont {K.}~\bibnamefont {Shlomi}}, \bibinfo
  {author} {\bibfnamefont {O.}~\bibnamefont {Shtempluck}}, \ and\ \bibinfo
  {author} {\bibfnamefont {E.}~\bibnamefont {Buks}},\ }\href@noop {} {\bibfield
   {journal} {\bibinfo  {journal} {arXiv:1210.7327}\ } (\bibinfo {year}
  {2012})}\BibitemShut {NoStop}%
\bibitem [{\citenamefont {Dorenbos}\ \emph {et~al.}(2008)\citenamefont
  {Dorenbos}, \citenamefont {{Reiger}}, \citenamefont {{Perinetti}},
  \citenamefont {{Zwiller}}, \citenamefont {{Zijlstra}},\ and\ \citenamefont
  {{Klapwijk}}}]{Dorenbos_131101}%
  \BibitemOpen
  \bibfield  {author} {\bibinfo {author} {\bibfnamefont {S.~N.}\ \bibnamefont
  {Dorenbos}}, \bibinfo {author} {\bibfnamefont {E.~M.}\ \bibnamefont
  {{Reiger}}}, \bibinfo {author} {\bibfnamefont {U.}~\bibnamefont
  {{Perinetti}}}, \bibinfo {author} {\bibfnamefont {V.}~\bibnamefont
  {{Zwiller}}}, \bibinfo {author} {\bibfnamefont {T.}~\bibnamefont
  {{Zijlstra}}}, \ and\ \bibinfo {author} {\bibfnamefont {T.~M.}\ \bibnamefont
  {{Klapwijk}}},\ }\href {\doibase 10.1063/1.2990646} {\bibfield  {journal}
  {\bibinfo  {journal} {Appl. Phys. Lett.}\ }\textbf {\bibinfo {volume} {93}},\
  \bibinfo {pages} {131101} (\bibinfo {year} {2008})}\BibitemShut {NoStop}%
\bibitem [{\citenamefont {Marsili}\ \emph {et~al.}(2012)\citenamefont
  {Marsili}, \citenamefont {Verma}, \citenamefont {Stern}, \citenamefont
  {Harrington}, \citenamefont {Lita}, \citenamefont {Gerrits}, \citenamefont
  {Vayshenker}, \citenamefont {Baek}, \citenamefont {Shaw}, \citenamefont
  {Mirin},\ and\ \citenamefont {Nam}}]{Marsili_arxiv}%
  \BibitemOpen
  \bibfield  {author} {\bibinfo {author} {\bibfnamefont {F.}~\bibnamefont
  {Marsili}}, \bibinfo {author} {\bibfnamefont {V.}~\bibnamefont {Verma}},
  \bibinfo {author} {\bibfnamefont {J.}~\bibnamefont {Stern}}, \bibinfo
  {author} {\bibfnamefont {S.}~\bibnamefont {Harrington}}, \bibinfo {author}
  {\bibfnamefont {A.}~\bibnamefont {Lita}}, \bibinfo {author} {\bibfnamefont
  {T.}~\bibnamefont {Gerrits}}, \bibinfo {author} {\bibfnamefont
  {I.}~\bibnamefont {Vayshenker}}, \bibinfo {author} {\bibfnamefont
  {B.}~\bibnamefont {Baek}}, \bibinfo {author} {\bibfnamefont {M.}~\bibnamefont
  {Shaw}}, \bibinfo {author} {\bibfnamefont {R.}~\bibnamefont {Mirin}}, \ and\
  \bibinfo {author} {\bibfnamefont {S.}~\bibnamefont {Nam}},\ }\href@noop {}
  {\bibfield  {journal} {\bibinfo  {journal} {arXiv:1209.5774}\ } (\bibinfo
  {year} {2012})}\BibitemShut {NoStop}%
\bibitem [{\citenamefont {Engel}\ \emph {et~al.}(2012)\citenamefont {Engel},
  \citenamefont {Inderbitzin}, \citenamefont {Schilling}, \citenamefont
  {Lusche}, \citenamefont {Semenov}, \citenamefont {H{\"u}bers}, \citenamefont
  {Henrich}, \citenamefont {Hofherr}, \citenamefont {Siegel} \emph
  {et~al.}}]{Engel_arxiv}%
  \BibitemOpen
  \bibfield  {author} {\bibinfo {author} {\bibfnamefont {A.}~\bibnamefont
  {Engel}}, \bibinfo {author} {\bibfnamefont {K.}~\bibnamefont {Inderbitzin}},
  \bibinfo {author} {\bibfnamefont {A.}~\bibnamefont {Schilling}}, \bibinfo
  {author} {\bibfnamefont {R.}~\bibnamefont {Lusche}}, \bibinfo {author}
  {\bibfnamefont {A.}~\bibnamefont {Semenov}}, \bibinfo {author} {\bibfnamefont
  {H.}~\bibnamefont {H{\"u}bers}}, \bibinfo {author} {\bibfnamefont
  {D.}~\bibnamefont {Henrich}}, \bibinfo {author} {\bibfnamefont
  {M.}~\bibnamefont {Hofherr}}, \bibinfo {author} {\bibfnamefont
  {M.}~\bibnamefont {Siegel}},  \emph {et~al.},\ }\href@noop {} {\bibfield
  {journal} {\bibinfo  {journal} {arXiv:1210.5395}\ } (\bibinfo {year}
  {2012})}\BibitemShut {NoStop}%
\end{thebibliography}%
\end{document}